\documentclass{aastex}          %% The manuscript based on AASTeX v5.x
\usepackage{spr-astr-addons}    %% mimicing ASTR journal style
\RequirePackage{color}
\newcommand{\etal}{{et~al.\null}}
\newcommand{\eg}{{e.g.,}}

\newcommand{\ie}{{i.e.,}}
\newcommand{\oiii}{[O~{\footnotesize III}]}
\newcommand{\nii}{[N~{\footnotesize II}]}
\begin{document}
%% Article title
%
\title{The Planetary Nebula Luminosity Function at the Dawn of Gaia}

%% Running heads
\shorttitle{The Planetary Nebula Luminosity Function}
\shortauthors{Ciardullo}

%% Author and Affilations
\author{Robin Ciardullo\altaffilmark{1}}
\affil{The Pennsylvania State University}
\email{rbc@astro.psu.edu} %% non-output

%% Alternate Affilations
\altaffiltext{1}{Institute for Gravitation and the Cosmos, The Pennsylvania
State University}

%% Abstract
\begin{abstract}
The \oiii\ $\lambda 5007$ Planetary Nebula Luminosity Function (PNLF) is
an excellent extragalactic standard candle.  In theory, the PNLF method
should not work at all, since the luminosities of the brightest planetary 
nebulae (PNe) should be highly sensitive to the age of their host stellar 
population.  Yet the method appears robust, as it consistently produces
$\lesssim 10\%$ distances to galaxies of all Hubble types, from the earliest 
ellipticals to the latest-type spirals and irregulars.  It is therefore 
uniquely suited for cross-checking the results of other techniques and 
finding small offsets between the Population~I and Population~II distance 
ladders.  We review the calibration of the method and show that the zero 
points provided by Cepheids and the Tip of the Red Giant Branch are in 
excellent agreement.  We then compare the results of the PNLF with those 
from Surface Brightness Fluctuation measurements, and show that, although 
both techniques agree in a relative sense, the latter method yields distances 
that are $\sim 15\%$ larger than those from the PNLF\null.  We trace 
this discrepancy back to the calibration galaxies and argue that, due to 
a small systematic error associated with internal reddening, the true 
distance scale likely falls between the extremes of the two methods.  
We also demonstrate how PNLF measurements in the early-type galaxies that 
have hosted Type~Ia supernovae can help calibrate the SN~Ia maximum 
magnitude-rate of decline relation.  Finally, we discuss how the results 
from space missions such as Kepler and Gaia can help our understanding of 
the PNLF phenomenon and improve our knowledge of the physics of local 
planetary nebulae. 

\end{abstract}

%% Keywords

\section{Introduction}\label{s:1}
\begin{figure}[ht]
\begin{center}
\includegraphics[scale=0.356]{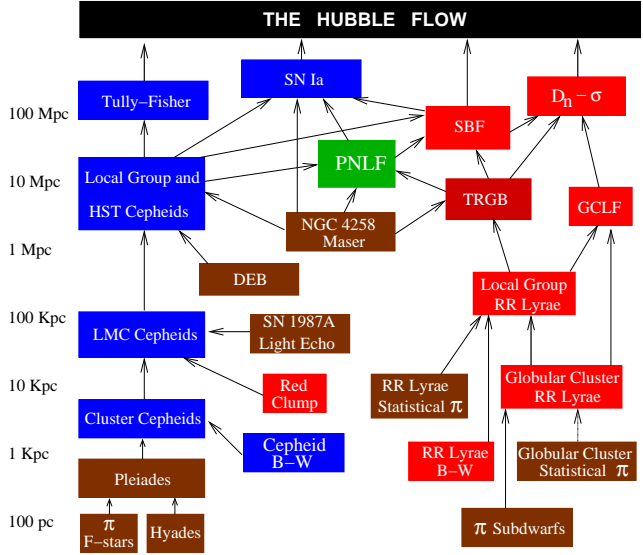}
\end{center}
\caption[]{\small The extragalactic distance ladder.  The blue boxes show 
techniques useful in star-forming galaxies, the red boxes give 
methods best suited for Pop~II systems, and the brown boxes represent 
geometric measurements.  Since the PNLF is equally effective in all
stellar populations, it is ideal for cross-checking the results of 
different methods.}
\label{ladder}
\end{figure}

The \oiii\ $\lambda 5007$ Planetary Nebulae Luminosity Function (PNLF) has 
been a reliable and precise extragalactic distance indicator for over 
$\sim 20$~years \citep{paper1, paper2, mudville}.  During this time, the
method has been applied to both elliptical \citep{paper5} and spiral 
\citep{paper11} galaxies, to galactic spheroids \citep{paper2, paper3, cena} 
and disks \citep{paper11}, and even to stars lost within the
intergalactic environment of rich clusters \citep{ipn3}.  In addition,
PNLF observations are relatively easy:  the method requires neither space-based
observations nor heroically long integrations, and the photometric
procedures needed to derive accurate distances are simple and straightforward.
As a result, the PNLF is an integral part of the extragalactic
distance ladder, and perhaps the best tool we have for examining systematic
differences between Population~I and Population~II distance methods
(see Figure~\ref{ladder}).

Of course, no technique is perfect.  The PNLF is not well-suited for small
galaxies which contain few planetary nebulae, nor can it reach to
$\sim 100$~Mpc, where the unperturbed Hubble Flow dominates.  For all
practical purposes, PNLF observations are limited to $\sim 20$~Mpc (but 
see \citet{coma} and \citet{hydra} for PN measurements in Coma and Hydra!).  
The technique also requires the use of a narrow-band \oiii\ $\lambda 5007$ 
filter (FWHM $\sim 50$~\AA), whose properties in the converging beam of a 
telescope are well known. In this age of fast, extremely wide-field imagers, 
this can be a severe limitation, especially for systems where the
5007~\AA\ line is redshifted out of the bandpass of a rest-frame
\oiii\ filter.   Finally, the PNLF cannot easily be calibrated in the Milky
Way (but see \citet{mendez} and \citet{kovacevic2} for attempts at a local 
calibration), and we are far away from having a theoretical understanding of 
the method.

\section{Why the PNLF Can't Work}
The universe constantly surprises us.  For example, in the 1980's, if one
were to ask whether 4-m telescopes could detect \oiii\ $\lambda 5007$
emission from planetary nebulae in the Virgo Cluster core, the answer would
have certainly been no.  The main-sequence turnoff of an elliptical galaxy is 
$\sim 1 M_{\odot}$, which, through the initial mass-final mass relation, means 
that the central stars of the PNe currently being produced have relatively 
small masses, \ie\ $\sim 0.52 M_{\odot}$ \citep{kalirai}.  It is well 
established that cores in this mass range cannot be more luminous than
$\sim 1000 \, L_{\odot}$ \citep{schon83, vw94}, and that less than $\sim 12\%$ 
of this flux can be reprocessed into the \oiii\ $\lambda 5007$ emission line
\citep{djv92, schonberner10}.  Thus, at maximum, the PNe inside the 
elliptical galaxies of Virgo should have \oiii\ monochromatic fluxes of 
$1.2 \times 10^{-17}$~ergs~cm$^{-2}$~s$^{-1}$ ($m_{5007} \sim 28.5$), well
below the threshold for night-long observations with a 4-m class telescope.

Of course, one might argue that real galaxies contain a mix of stellar
populations and that there will always be some PNe produced by higher mass 
stars.  But this claim poses another problem.   As elegantly pointed out 
by \citet{renzini}, the number of PNe produced by any single-aged stellar
population is linearly proportional to the luminosity of that population, \ie\
\begin{equation}
N(PN) = B \, \cdot \, t \, \cdot \, L 
\end{equation}
where $t \sim 500$~yr is the lifetime of a PN in its \oiii-bright phase
\citep{marigo, schonberner07}, $L$ is the total integrated (bolometric) 
luminosity of the population in question, and $B$ is the population's 
luminosity specific stellar evolutionary flux.  Remarkably, $B$ does not 
depend on age, metallicity, or initial mass function:  to within $\sim 10\%$, 
all populations older than $\sim 1$~Gyr have 
$B \sim 2 \times 10^{-11}$~stars~yr$^{-1}$~$L_{\odot}^{-1}$ \citep{renzini}.  
Consequently, in order for a Virgo elliptical to have a sizeable number
of \oiii-bright PNe, $\sim 10\%$ of its stars must be no older than
$\sim 1$~Gyr.  Such a result has long been ruled out via integrated light 
spectroscopy \citep[\eg][]{vazdekis, trager00}.

Furthermore, even if the PNe were detected, there is absolutely no reason
to expect the bright end of the PNLF to be a standard candle.
From stellar evolution theory, the \oiii\ 
$\lambda 5007$ magnitude of the brightest PNe produced by a simple (single 
age) stellar population should change with time, following
\begin{eqnarray}
\frac{d M_{5007}}{dt} &=&
\left( \frac{d M_{5007}}{d\log L_{5007}} \right) \cdot
\left( \frac{d \log L_{5007}} {d \log L_*} \right) \cdot \nonumber \\
& & \ \ \ \ \ \ \left( \frac{d \log L_*}{dm_f} \right) \cdot
\left( \frac{d m_f}{d m_t} \right) \cdot 
\left( \frac{d m_t}{dt} \right) \nonumber \\
\noalign{\vskip5pt}
&=&(-2.5) \cdot (\sim 1) \cdot (\sim 7) \cdot (0.11) \cdot (0.8 \, t^{-1.4}) 
\nonumber \\
\noalign{\vskip5pt}
&=& \sim -1.5 \, t^{-1.4} 
\label{eq:deriv}
\end{eqnarray}
where $L_*$ is the luminosity of the central star, $m_f$ is the mass
of the PN core, $m_t$ is the turnoff mass of the stellar population, and $t$
is the age of the population (measured in Gyr).  All these derivatives are 
reasonably well known via our knowledge of main sequence and post-AGB stellar 
evolution \citep[\eg][]{iben, vw94}, the initial mass-final mass relation 
\citep[\eg][]{kalirai, dobbie}, and nebular physics \citep[\eg][]{ferland, 
perinotto}.  (The most uncertain term involves the reprocessing of stellar 
luminosity into line emission at 5007~\AA, but models such as those by 
\citet{schonberner10} suggest that this derivative is a slowly varying
quantity.)  For old stellar systems ($t \sim 10$~Gyr), 
equation~(\ref{eq:deriv}) implies that the PNLF cutoff should fade by 
$\sim 0.1$~mag per Gyr, and over a timescale of $\sim 10$~Gyr, the decline
in brightness should be more than $\sim 4$~mag \citep{marigo, schonberner07}.  
In this age of precision cosmology, such a strong dependence would have been 
certainly been seen.

\begin{figure}[h]
\begin{center}
\includegraphics[scale=0.471]{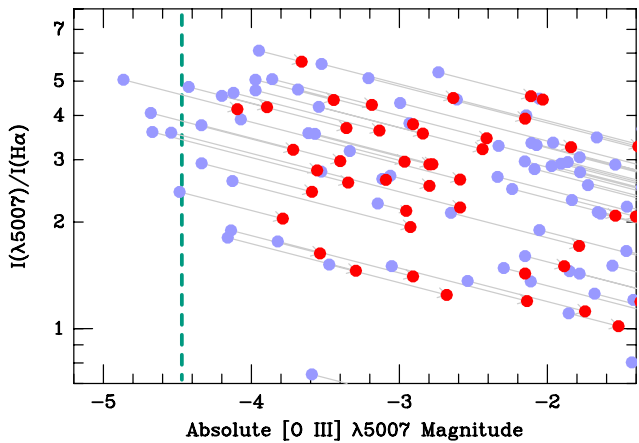}
\end{center}
\caption[]{The \oiii\ $\lambda 5007$ to H$\alpha$+\nii\ line ratio 
plotted against \oiii\ absolute magnitude for a complete sample of
PNe in the LMC \citep{mash}.   The light blue points show the intrinsic line 
strengths; the red points illustrate the line intensities that are observed.
The dotted green line displays the observed value of the PNLF cutoff, $M^*$.  
The plot suggests that $M^*$ is largely defined by the action of 
circumstellar dust.}
\label{mash}
\end{figure}

Finally, an examination of planetary nebula morphologies in the Milky Way 
reveals that most bright PNe are asymmetric, and have position-dependent dust 
opacities within their circumstellar envelopes \citep[\eg][]{ueta, siodmiak}.  
The effect this dust has on the emergent \oiii\ $\lambda 5007$ emission is 
significant: although each PN is different, the average \oiii-bright planetary 
nebula is self-extincted by more than 0.6~mag at 5007~\AA\ \citep{herrmann2,
mash}.  In fact, as Figure~\ref{mash} shows, the bright-end cutoff of 
the PNLF is largely defined by the effects of sight-line dependent 
circumstellar extinction.  At face value, this fact alone would seem to 
preclude the PNLF from being an effective distance indicator.

\section{But the PNLF Does Work}
\begin{figure}[ht]
\begin{center}
\includegraphics[scale=0.433]{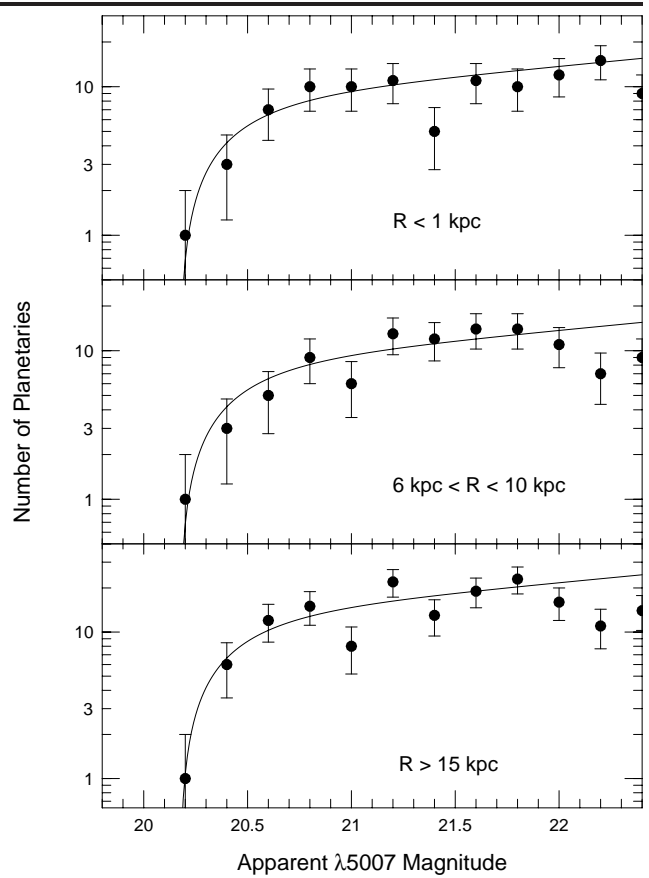}
\end{center}
\caption[]{\small The observed planetary nebula luminosity functions for 
samples of M31 PNe projected at three different galactocentric radii.  The 
curves show the best-fitting empirical law.  Although the stellar
populations of M31's inner bulge ($R < 1$~kpc), inner disk ($6~{\rm kpc} < 
R < 10$~kpc), and outer disk/halo ($R > 15$~kpc) are vastly different, 
the derived PNLF distances to each region are 
consisted to within $\sim 0.05$~mag.}
\label{m31}
\end{figure}

Despite these arguments, the PNLF is a precise extragalactic standard
candle, capable of generating distances to better than $\sim 10\%$ in
clusters as far away as Virgo and Fornax.  To demonstrate this, one need 
only look at our Local Group neighbor, M31.  As illustrated in 
Figure~\ref{m31}, the PNLF of M31's bulge ($R < 1$~kpc) has the same 
bright-end cutoff (defined via the absolute magnitude $M^*$) as that of 
the system's inner disk ($6 < R < 10$~kpc) and outer disk/halo ($R > 15$~kpc). 
Given the range of stellar population being observed, this invariance is 
remarkable.

M31 is not the only place where the PNLF has been tested.  PN studies 
in galaxies as diverse as M33 \citep{m33}, NGC~5128 \citep{cena}, 
and NGC~4494 \citep{paper10}, have been unable to detect any change
in $M^*$ with galactocentric radius.  Similarly, observations
of galaxies in groups and clusters, such as Triangulum \citep{paper7}, 
the M81 Group \citep{paper3, paper12}, Leo~I \citep{paper4, paper11, paper12}, 
Virgo \citep{paper5, pn_sn}, and Fornax \citep{paper9, n1344, pn_sn} (almost) 
always place all the galaxies comfortably within the typical group diameter 
of $\sim 1$~Mpc.  (The lone exception is in Virgo, where the PNLF clearly 
resolves the M84/M86 system which is known to be infalling into Virgo from 
behind.) The multitude of internal tests place strong constraints on the 
types and amplitudes of any systematic errors that might be associated with 
the technique.  The tests also demonstrate that the PNLF is capable of 
generating relative extragalactic distances to just a few percent in a 
variety of galactic environments.

\begin{figure}[t]
\begin{center}
\includegraphics[scale=0.455]{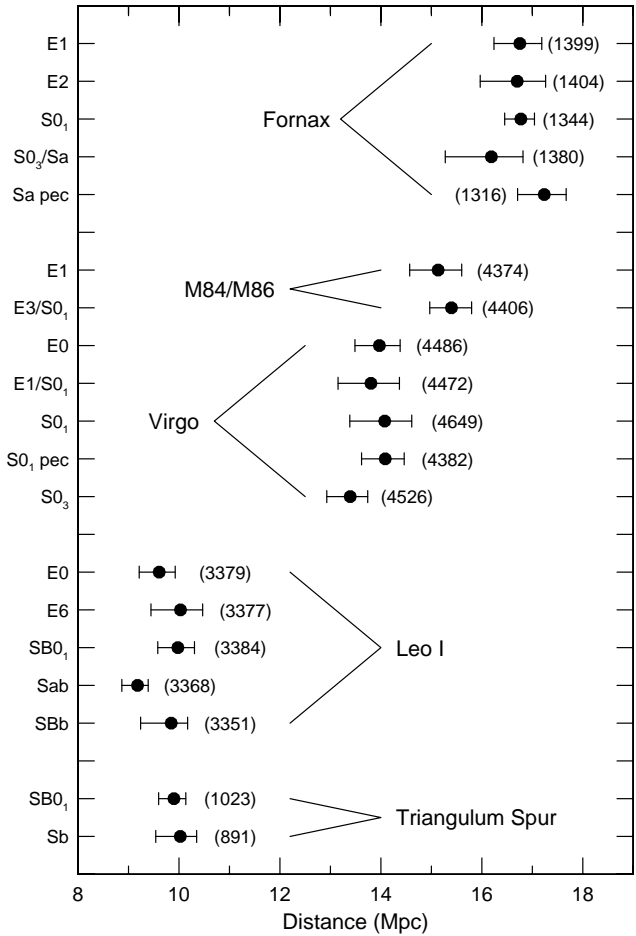}
\end{center}
\caption[]{\small PNLF distance measurements in the Triangulum Spur, the
Leo~I Group, the Virgo Cluster, and the Fornax Cluster.  
The M84/M86 system is also marked:  although projected onto
the Virgo Cluster core, this group is actually falling into Virgo
from behind.   The Hubble types are from \citet{rsa}.  The
diversity of Hubble types illustrates the robustness of the PNLF method.}
\label{distances}
\end{figure}

\section{The Calibration of the PNLF}

There is no theoretical explanation for the constancy of the PNLF, and
reliable distances to Galactic planetary nebulae are few and far between
\citep{phillips, harris}.  Hence, the only way to obtain an absolute 
measurement of $M^*$ and look for external errors in the PNLF method is to 
survey extragalactic systems that have distances known from other methods.
\citet{paper12} did this via PN observations inside
13 galaxies with Cepheid distances derived by the {\sl HST\/} Key Project 
\citep{keyfinal}.  By adopting the ``universal'' PNLF first proposed by 
\citet{paper2},
\begin{equation}
N(M) \propto e^{0.307 M} \{ 1 - e^{3 (M^* - M)} \}
\label{eq:pnlf}
\end{equation}
the authors obtained a value of $M^* = -4.47$, where 
\begin{equation}
M_{5007} = -2.5 \log F_{5007} - 13.74
\label{eq:mstar}
\end{equation}
and $F_{5007}$ is the \oiii\ $\lambda 5007$ flux in 
ergs~cm$^{-2}$~s$^{-1}$.  

Remarkably, despite $\sim 20$~years of observations, non-parametric statistical
tests such as Kolmogorov-Smirnov and Anderson-Darling still have not found any
reason to reject the shape defined by this simple analytic function (but see
the slitless spectroscopy of \citet{sambhus} for a curious result in the
elliptical galaxy NGC~4697).  Of course, this does not mean that other 
formulations of the PNLF
are not possible.  For example, the recent distance determinations to 
NGC~4697 \citep{n4697}, NGC~1344 \citep{n1344}, M82 \citep{m82}, NGC~821 
\citep{n821}, and NGC~4649 \citep{n4649} all use a numerical form of the PNLF 
derived by combining models of post-AGB evolution with empirical constraints 
on the excitation of \oiii $\lambda 5007$ in Galactic and Magellanic Cloud 
planetaries \citep{men-soff}.  These measurements are self-consistent within
themselves, and usually produce distances that are very similar to those found 
from equation~(\ref{eq:pnlf}).  However, because the functions are different,
there may be small ($\lesssim 0.1$~mag) systematic offsets between the
distances derived in this way and those inferred from the purely empirical 
PNLF\null.  The quoted uncertainties of the individual distances may also 
have slightly different meanings.

\begin{figure*}[t]
\begin{center}
\includegraphics[scale=0.846]{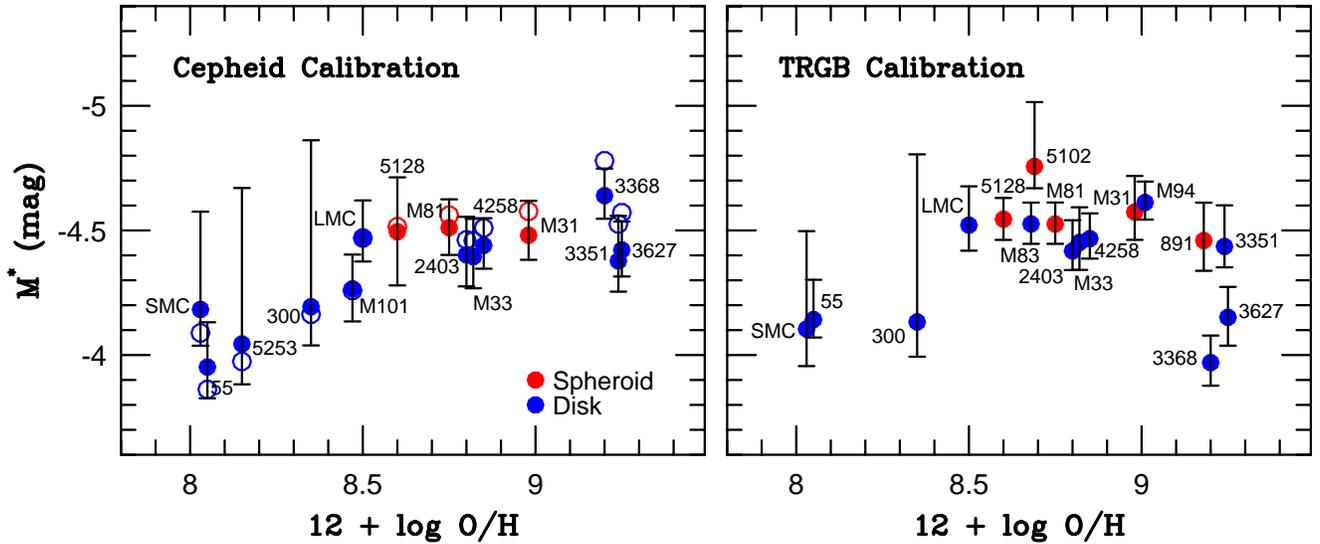}
\end{center}
\caption[]{\small Measurements of $M^*$ derived using distance estimates from
Cepheids (left panel) and the Tip of the Red Giant Branch (right panel).
Most of the Cepheid distances come from \citet{keyfinal} and assume no
metallicity correction.  The open circles show the same calibration with an
assumed value of $\gamma = -0.2$ for the dependence of the Cepheid 
period-luminosity relation on metallicity.  Such a correction increases $M^*$ 
by $\sim 0.07$~mag, while increasing the dispersion only slightly. The TRGB 
distances come principally from \citet{dalcanton09}, \citet{jacobs09}, and 
\citet{hislop11}.}
\label{comp}
\end{figure*}

Since 2002, a few of the PNLF and Cepheid measurements have improved,
and the number of galaxies studied by both methods has increased to 16\null.
But qualitatively, little has changed.  As the left-hand panel of
Figure~\ref{comp} illustrates, all stellar populations more metal-rich
than the LMC appear to have the same value for the PNLF cutoff, 
$M^* \sim -4.5$.  In the smaller, metal-poor systems, $M^*$ does fade, in 
general agreement with the nebular model predictions of \citet{djv92} and 
\citet{schonberner10}.  However, since these low-mass systems have few
PNe and poorly defined PN luminosity functions, this systematic shift in
$M^*$ is not an important limitation of the method.

We do note that the best-fit value of $M^*$ can be changed slightly,
depending on how one models the response of the Cepheid period-luminosity
relation to metallicity.  The solid points displayed in the left-hand 
panel of Figure~\ref{comp} adopt the Cepheid distances of \citet{keyfinal}
and assume that the period-luminosity relation for Cepheids has no dependence 
on metallicity.  If we exclude the small, low-metallicity galaxies (\ie\
objects with $12 + \log {\rm O/H} < 8.45$), these data imply a most likely value
for the PNLF cutoff of $M^* = -4.46 \pm 0.05$ (standard deviation of the 
mean), and an external scatter ($\sigma = 0.16$~mag) that is fully consistent 
with the internal errors of the measurements.  Alternatively, if the
\citet{keyfinal} period-luminosity data are used with their suggested 
metallicity dependence ($\gamma = -0.2$), then the most likely value of
$M^*$ brightens to $M^* = -4.53 \pm 0.04$, while the external scatter 
of the individual $M^*$ measurements increases only marginally to 0.18~mag.  
Finally, an extreme value of $M^* = -4.68 \pm 0.09$ can be obtained if the 
Cepheid distances and metallicities of \citet{saha} are used in the analysis.
Under this assumption, however, the internal errors of the individual PNLF 
measurements are no longer consistent with the observed galaxy-to-galaxy
scatter, which explodes to almost 0.3~mag.

Rather than relying solely on the Cepheid period-luminosity relation, we
can take another route to the PNLF zero point by using distances 
inferred from measurements of the Tip of the Red Giant Branch 
(TRGB\null).  Thanks mostly to {\sl HST,} 18 PNLF galaxies of varying Hubble
types now have reliable distances from this technique \citep{dalcanton09, 
jacobs09, hislop11}.  The values of $M^*$ based on these measurements are 
shown in the right-hand panel of Figure~\ref{comp}.  If we exclude the results 
from the two distant systems (NGC~3368 and 3627) where the TRGB and Cepheid 
distances are in conflict, the plot looks remarkably similar to that obtained 
from the Cepheids.  As before, there is evidence for a decrease in $M^*$ in 
low-luminosity, low-metallicity systems.  However, in systems more metal-rich 
than the LMC, $M^*$ again has a best-fit value of $M^* = -4.53 \pm 0.06$ and 
there is no evidence for a metallicity dependence.  The fact that this number, 
with its small dispersion, agrees with the calibration derived from Cepheids 
supports the argument that the zero point of the PNLF system is secure at 
the $\sim 5\%$ level.

\section{The PNLF and Surface Brightness Fluctuations}
Both the Cepheid and TRGB methods establish the zero point of the 
PNLF distance scale to $\sim 5\%$.  We can now use this fact to
investigate the calibration of other methods, and search for systematic
errors in the distance ladder.  For example, the Surface Brightness 
Fluctuation (SBF) method is a popular way of obtaining
high-precision distances to galaxies with smooth luminosity profiles,
\ie\ elliptical and lenticular systems \citep{sbf1, tonry01}.  The
technique is efficient, well calibrated \citep{tonry01, mei05,
blakeslee10}, and has a solid theoretical foundation \citep[\eg][]{liu+00,
blakeslee+01, mfa}.  The most direct way of obtaining the zero point of this 
method is to use the seven Cepheid galaxies with bulge populations large enough
to be measured with the SBF technique.  If the zero point derived from
these galaxies is accurate, then a comparison of PNLF and SBF distances 
should show good agreement, with a mean near zero, and a scatter that is 
representative of the internal errors of the methods.

\begin{figure}[t!]
\begin{center}
\includegraphics[scale=0.490]{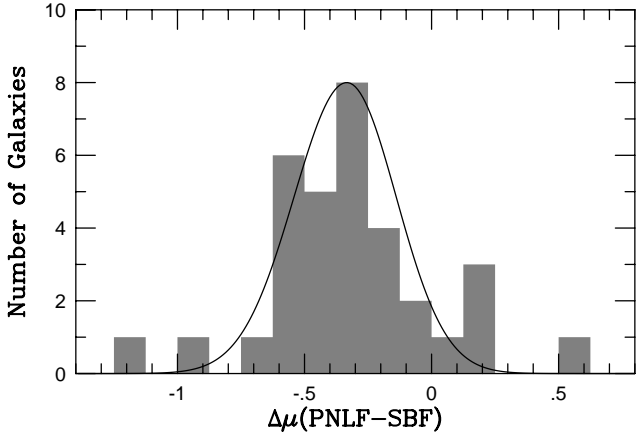}
\end{center}
\caption[]{A histogram of the differences between the PNLF and SBF
distance moduli for 33 galaxies measured by both techniques.  The two worst
outliers are the edge-on galaxies NGC~4565 ($\Delta\mu = -1.14$) and
NGC~891 ($\Delta\mu = +0.56$).  NGC~4278 is also an outlier ($\Delta\mu =
-0.97$).  The curve represents the expected dispersion of the data.
The figure demonstrates that, although the internal and external
errors of the methods agree, the absolute scales defined the two techniques
are in conflict.
}
\label{sbfhist}
\end{figure}

Figure~\ref{sbfhist} displays the difference in distance moduli for
the set of 33 PNLF galaxies with SBF $I$-band measurements 
\citep{tonry01, blakeslee10}.   It is immediately obvious that the 
systematic offset between the two methods is not zero, nor is it close
to zero.  The SBF distance scale is, in the median, $\Delta \mu = 0.33$
larger than PNLF distance scale.  Moreover, this property is not
restricted to the groundbased $I$-band dataset.  Twelve PNLF galaxies have
high-quality SBF measurements in the $z$-band from observations with the
{\sl ACS\/} on {\sl HST\/} \citep{blakeslee09}; the offset between their PNLF 
and SBF distance moduli is $\Delta\mu = 0.42$.  Similarly, 16 galaxies have 
PNLF and SBF $H$-band data from {\sl HST's\/} NICMOS camera 
\citep{jensen}.  The median offset between these two datasets is 
$\Delta\mu = 0.29$.  Clearly, there is a problem with the zero point of one 
(or both) of the techniques.

\begin{figure*}[t!]
\begin{center}
\includegraphics[scale=0.919]{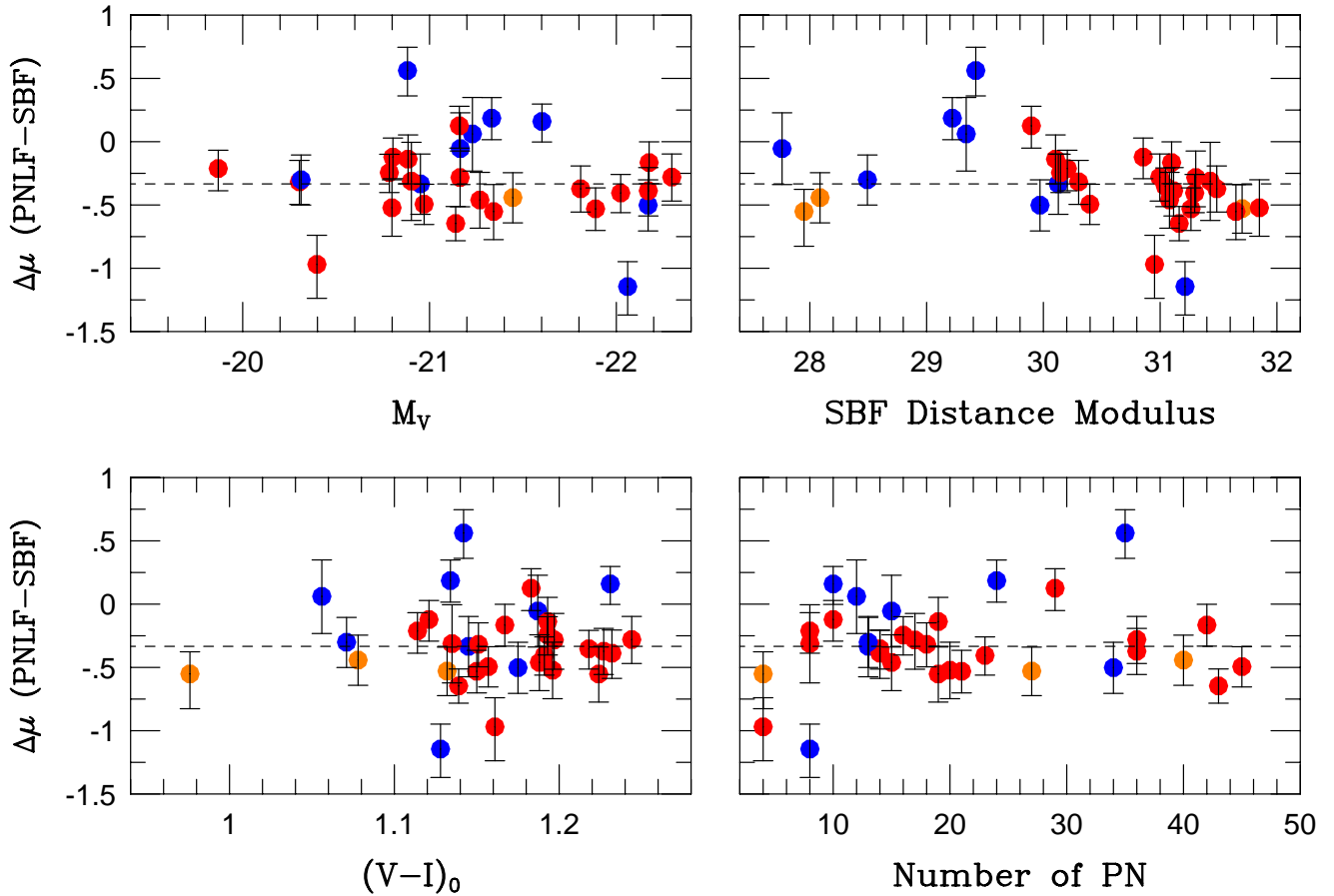}
\end{center}
\caption[]{The difference between SBF and PNLF distance moduli plotted
against galactic absolute magnitude, distance, color, and number of PNe 
within 0.5~mag of $M^*$.  The red points represent elliptical and
lenticular galaxies, the orange points are gas-rich lenticulars and/or 
recent mergers, and the blue points are normal spiral
galaxies with SBF measurements in their bulge.  The correlation with 
SBF distance modulus, though significant ($P \sim 0.01$), disappears
if spiral galaxies are excluded from the analysis.   Note that the 
scatter for spiral galaxies is much larger than that for the early-type
systems.}
\label{sbf_scatter}
\end{figure*}

The fact that the offset is due to a zero point error, rather than a systematic
trend with stellar population or distance can be shown in two ways.  First, 
the solid curve in Figure~\ref{sbfhist} is not a fit to the data:  it is
instead, the {\it expected\/} scatter in the measurements, as
determined by propagating the uncertainties associated with the 
individual PNLF and SBF distances and Galactic reddening.
The agreement between the curve and data proves that the quoted 
uncertainties of both methods are reasonable, and that there is little
room for additional random errors associated with the measurements.
Similarly, if we plot the individual differences in distance moduli against
against external parameters such as galactic absolute magnitude, color,
distance modulus, or number of bright PNe (within 0.5~mag of $M^*$),
we find only one significant trend.  As Figure~\ref{sbf_scatter} illustrates, 
the offset between the PNLF and SBF correlates slightly with SBF distance
modulus, in the sense that distant galaxies have smaller PNLF distances
than expected from their fluctuation magnitude.  \citet{ferrarese} interpreted 
this trend as being due to a systematic error in the PNLF that only effects
galaxies with distances greater than 10~Mpc.  At the time, such an error
could plausibly have been due to the existence of foreground intracluster 
stars in the clusters of Virgo and Fornax.   However, in the past few
years, this explanation has become untenable.  Not only is the hypothesis
at odds with the information provided by deep Virgo Cluster surface photometry
\citep{mihos+05, mihos+09}, but additional data has shown that the distance
discrepancy is not restricted to galaxies in the centers of clusters.  Objects 
such as the isolated elliptical NGC~821 and the x-ray faint elliptical
NGC~4697 (in the Virgo southern extension), display the same curious 
offset.

In fact, a closer look at Figure~\ref{sbf_scatter} shows that the apparent
correlation with SBF distance modulus is not driven so much by the 
measurements of the most distant systems, as it is by the behavior of the 
relatively nearby spiral galaxies.  If these later-type objects
are excluded from the analysis, the correlation with distance disappears.
Moreover, except for the systematic offset between the two techniques, the 
only consistent pattern visible in Figure~\ref{sbf_scatter} involves 
the dispersion: the scatter between the PNLF and SBF distance estimates
for spiral galaxies is much larger than what is seen for elliptical 
and lenticular systems.

%\begin{figure*}[ht!]
%\begin{center}
%\includegraphics[scale=0.818]{Ciardullo_fig8.eps}
%\end{center}
%\caption[]{The difference between SBF, Cepheid, and TRGB distance moduli 
%plotted against the color of the underlying stellar population.  The
%Cepheid distances come from \citet{keyfinal} and assume no metallicity
%dependence.  The data suggest
%that the SBF zero point may be $\sim 0.2$~mag too large.
%}
%\label{sbf_comp}
%\end{figure*}

So where does the offset come from?  As Figure~\ref{comp} demonstrates, 
the PNLF zero point is well calibrated by both Cepheid and TRGB
measurements; while a small increase in the distance scale may be possible,
any increment greater than $\sim 0.1$~mag is excluded by the data.  
Likewise, an examination of the SBF zero point calibration shows little 
room for error \citep[\eg][]{ajhar, ferrarese}.  The median offset for the 
seven galaxies with both SBF and Cepheid distance determinations is 
$\Delta\mu = -0.10$ if the period-luminosity relation is independent of metal 
abundance, or $+0.08$ if $\gamma = -0.2$.  It should be noted, however, that 
the uncertainties associated with several of the individual fluctuation 
magnitudes used in this calibration are large, and a similar analysis
with the 12 SBF galaxies with TRGB measurements yields an offset of 
$\Delta\mu = -0.17$.   Thus, while there is some evidence to support a 
modification in the SBF zero point, any downward adjustment to its distance 
scale must be slight, $\lesssim 0.15$~mag.  Another explanation is therefore 
needed to reconcile the PNLF-SBF discrepancy.

The answer to this paradox is internal extinction.  
To calibrate an extragalactic standard candle,
one needs to measure the apparent brightness of the candle, $m$, and adopt
values for the distance modulus to the galaxy, $\mu$, and the intervening
extinction, $E(B-V)$.  In other words,
\begin{equation}
M = m - \mu - R_{\lambda} \, E(B-V)
\label{eq:dmod}
\end{equation}
Here, $M$ is the derived absolute magnitude of the standard candle in
question and $R_{\lambda}$ is the ratio of total to differential extinction
at the wavelength of interest.  (For \oiii\ $\lambda 5007$ observations
of planetary nebulae, the extinction law of \citet{ccm} gives 
$R_{\lambda} = 3.5$.)  For most methods, including the
PNLF, if the reddening to a galaxy is underestimated, then the brightness
of the standard candle will be underestimated, and the distance scale implied
by the observations will be underestimated.  However, in the case of the 
SBF technique, there is a non-negligible color term, which causes redder
stellar populations to have dimmer fluctuation magnitudes.  Consequently, the 
absolute fluctuation magnitude, $\overline M$, is given by
\begin{equation}
\overline M_I = C + a (V-I)_0
\label{eq:sbf}
\end{equation}
where $a$ is the slope of the color dependence.
The zero-point of the SBF distance scale, represented by the constant $C$, is 
therefore defined through
\begin{equation}
C = \overline m_I - \mu - a (V-I)_{\rm obs} + (a \, R_V - (a+1) \, R_I) \,
E(B-V)
\label{eq:sbf_c}
\end{equation}
For the $I$-band SBF measurements, $a = 4.5$, $R_V = 3.04$, and $R_I = 1.88$,
so 
\begin{equation}
C = \overline m_I - \mu + 3.34 E(B-V) 
\label{eq:offset}
\end{equation}
(For the {\sl HST\/} $z$-band photometry, the coefficient for the reddening
term is +1.71, while for the IR fluctuation data, the value is +5.35.) 
So, while the response of SBF measurements to reddening has roughly the 
same amplitude as that of the PNLF, it has the opposite sign!  Even if both 
analyses were performed perfectly, our imperfect knowledge of 
foreground extinction would cause the two distance methods to produce 
different results. 

Of course, simple random uncertainties in the estimates of foreground 
reddening, such as those associated with the extinction maps of 
\citet{schlegel}, should not produce a systematic error between the two
distance scales.  However, the calibrators for the SBF and PNLF techniques
are mostly mid-type spiral galaxies, while the vast majority of the 
methods' program objects are elliptical and lenticular systems.  This 
population difference {\it can\/} create a systematic bias between the two
systems.  If the calibrator galaxies have, on average, as little as 
$E(B-V) \sim 0.02$ more internal reddening than the early-type systems that 
are the methods' main targets, the result will be a 0.15~mag offset between the 
PNLF and SBF distances, in exactly the direction that is observed.  This 
seems quite plausible, especially since the bulges of most mid-type spirals
exhibit more evidence for dust than do the interiors of normal 
elliptical galaxies \citep{windhorst}.  In fact, given the extremely strong
amplitude of the extinction dependence, it would be surprising if there 
weren't an offset!  

From the above arguments, it seems clear that the offset between the PNLF
and SBF distance scales is likely the result of several small effects:
a slight ($\lesssim 0.1$~mag) underestimate in the brightness of the PNLF
cutoff, a slight ($\lesssim 0.2$~mag) overestimate in the zero point of 
the SBF method, and a small shift in both the PNLF and SBF caused by the 
systematically larger dust content of spiral bulges.  Without the 
cross-checking ability of the PNLF, such subtle effects could easily 
go unnoticed.

\section{The PNLF and SN Ia}
Over 50 galaxies currently have robust PNLF distance measurements.  
These galaxies have hosted $\sim 50$ known supernovae in the past
century, and about half of these objects have well-observed light curves.  
Included in this sample of well-studied supernovae are 6 SN II-P, 2 SN II-L, 
1 SN IIb, 1 SN IIn, 1 SN Ib, 1 SN Ic, and 13 SN Ia (see Table~\ref{tab:sn}). 
The latter can be used to derive an estimate of $H_0$.

\begin{center}
\begin{table}[h]
\caption{SN Ia in PNLF Galaxies}
\begin{tabular}{@{}lll}
\hline
SN Ia &Host &Comment \\
\hline
1972E    &NGC 5253   &Photoelectric Photometry\\
1980N    &NGC 1316   &Photoelectric Photometry\\
1981D    &NGC 1316   &Photoelectric Photometry\\
1986G    &NGC 5128   &High Internal Extinction\\
1989B    &NGC 3627   &   \\
1991bg   &NGC 4374   &Subluminous \\
1992A    &NGC 1380 \\
1994D    &NGC 4526 \\
1998bu   &NGC 3368 \\
2006dd   &NGC 1316 \\
2006mr   &NGC 1316  &Subluminous \\
2007on   &NGC 1404 \\
2011fe   &NGC 5457 \\
\hline
\end{tabular}
\label{tab:sn}
\end{table}
\end{center}

\begin{figure}[ht]
\begin{center}
\includegraphics[scale=0.462, angle=0]{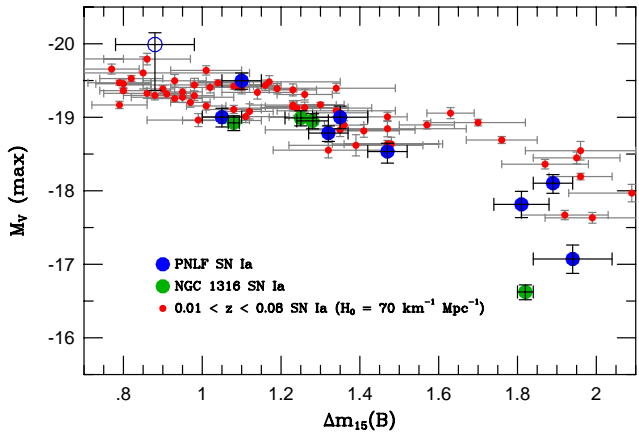}
\end{center}
\caption[]{\small The absolute $V$ magnitudes of 13 SN~Ia located within
galaxies with PNLF distances, plotted against the amount of fading which
occurred in the first 15 days after maximum.  For comparison,
a sample of supernovae in the redshift range $0.01 < z < 0.08$ 
is also shown, under the assumption of $H_0 = 70$~km~s$^{-1}$~Mpc$^{-1}$.
Except for SN 1972E (in NGC~5253, shown as an open circle), all the 
supernovae are from the past 30 years. 
}
\label{sn_comp}
\end{figure}

Figure~\ref{sn_comp} displays the $V$-band maximum magnitude-rate of decline 
relation for a set of well-observed supernovae in the Hubble flow, \ie\  in
the redshift range $0.01 < z < 0.08$ \citep{hicken09}, under the assumption 
that $H_0 = 70$~km~s$^{-1}$~Mpc$^{-1}$.  Also plotted are the maximum absolute
magnitudes of 13 well-observed local SN~Ia, using distances derived from the
PNLF\null.  The agreement between the two sets of data is surprising good,
especially when one considers that several of the supernovae are from
the era prior to CCD measurements and another is affected by high internal
extinction.  This consistency demonstrates the viability of the PNLF as a SN~Ia 
calibrator, and its utility.  Over half the local supernovae plotted 
in the figure occurred in early-type galaxies with no significant Population~I 
component.  Cepheids cannot be used to measure the distances to these objects.
In order to study the brightnesses of SN~Ia across galaxy types, one needs
a method such as the PNLF, as it is the most direct way to link measurements 
within Population~I and Population~II stellar systems.

\section{Prospects for the Future}
The greatest impediment to improving the PNLF distance scale is the lack of 
an adequate theory to explain why the technique is so robust.  The constancy 
of the PNLF cutoff cannot be explained by single-star stellar evolution:
the energy emitted by an $M^*$ planetary nebula demands that the PN has a
core mass of at least $\sim 0.6 M_{\odot}$ \citep{vw94}, and old stellar
populations simply cannot produce such high-mass objects.  The progenitors
of $\sim 0.6 M_{\odot}$ white dwarfs are main-sequence stars with masses
$M \sim 2 M_{\odot}$ \citep{kalirai}, and stars such as these
have very short main sequence lifetimes \citep[$\tau \lesssim 1$~Gyr;][]{iben}.
While early-type galaxies may contain a small number of intermediate-age
stars \citep{kaviraj}, this population is not nearly large enough to 
produce the number of PNe actually observed.  Consequently, in order
to explain the PNLF of elliptical galaxies, a more exotic channel of 
stellar evolution must be considered.

Binary evolution is the likely solution to this problem.  There is now
considerable evidence to suggest that a significant fraction of PNe form 
via binary star interactions \citep{soker, moe}.  Indeed, \citet{bs-pn} has 
proposed that an important channel for the creation of an \oiii-bright PN
involves conservative mass transfer whilst on the main sequence
\citep{mccrea}, and the subsequent creation of a blue straggler star.  Though
this hypothesis is difficult to test, clues as to its viability may soon be
provided by astrometric and photometric space missions.  \citet{clarkson} have
shown that space-based astrometry of the Galactic bulge can provide a census of
blue stragglers in a collisionless stellar environment with a well-studied
PN population \citep{pottasch, kovacevic1}.  By comparing the PN and blue
straggler number densities to their expected lifetimes 
\citep[\eg][]{schonberner10, lombardi}, one can test whether the
stellar merger hypothesis is self-consistent.  Similarly, if blue straggler
evolution is an important channel for populating the bright end of the 
planetary nebula luminosity function, it must also be a significant 
contributor to the A-star population of the solar neighborhood.  
Asteroseismological studies, such as those being performed with the Kepler 
satellite \citep{balona}, may be able to disentangle the two evolutionary 
scenarios, thereby shedding light on the creation rate of coalesced objects.

Another advance that we can look forward to is an improved distance
scale to PNe in the Milky Way.   Most Galactic planetary nebulae have 
distance estimates based on such dubious methods as the amount of foreground 
extinction \citep{giammanco} or the statistical correlation between nebular 
emission and nebular size \citep[\eg][]{shklovskii, vdsteene}.  Factor of 
$\sim 2$ errors in these estimates are not uncommon, even for bright, 
well-observed objects, and as a result, obtaining a local calibration of 
the PNLF cutoff is extremely difficult \citep{mendez, kovacevic2}.  This 
will soon change, however, as Gaia will expand the number of PNe with reliable 
parallax measurements from $\sim 20$ \citep{harris} to more than $\sim 200$. 
This should facilitate dozens of studies of the local PN population, 
including its \oiii\ luminosity function.

Unfortunately, a Gaia-based luminosity function, by itself, will probably not 
enable us to significantly improve the calibration of the PNLF\null.  As 
Figure~\ref{mash} illustrates, $M^*$ is highly dependent on the amount of
extinction produced by circumstellar dust, and, despite the 
extraordinary progress made in understanding the physics of the planetary 
nebula phenomenon \citep{schonberner10}, this component of the problem has 
yet to be modeled.   This is major issue, since observations of Galactic PNe 
are affected not only by the actions of circumstellar dust, but by foreground 
extinction as well.  The problem of disentangling the two reddening
components may ultimately limit any attempt to calibrate the planetary
nebula luminosity function in the Milky Way.

\section{Conclusion}
The PNLF is neither a primary distance indicator nor a method that
can reach out of the Local Supercluster into the Hubble Flow.  It
has no theoretical basis, and next generation missions such as 
Gaia will not substantially improve its calibration.  Yet the 
technique remains an important part of the extragalactic distance ladder:
it is, perhaps, the best tool we have for linking the Population~I and
Population~II distance scales, and identifying systematic offsets between the
different methods.   Moreover, it is still a relatively efficient technique
that does not require space-based observations or extremely long
exposure times.  Since PNLF distances are often produced as a by-product
of kinematic studies \citep[\ie][]{herrmann1, n821}, the method is
likely to be producing distance measurements for years to come.

\acknowledgments
We would like to thank the meeting's organizers for supporting this
review.  This work was supported by NSF grant AST 06-07416.

\end{document}